\documentclass[pra, twocolumn, floatfix, superscriptaddress, longbibliography]{revtex4-1}

\usepackage{amsmath, amssymb, bbm, bbold}
\usepackage{graphicx}
\usepackage{color}

\newcommand{\ket}[1]{{| #1 \rangle}}
\newcommand{\bra}[1]{{\langle #1 |}}
\newcommand{\ee}{{\rm e}}
\newcommand{\ii}{{\it i}}
\newcommand{\dd}{{\rm d}}
\newcommand{\Tr}{{\rm Tr}}
\newcommand{\sinc}{{\rm sinc}}

\begin{document}

\title{Disorder-dressed quantum evolution}

\author{Clemens Gneiting}
\email{clemens.gneiting@riken.jp}
\affiliation{Theoretical Quantum Physics Laboratory, RIKEN Cluster for Pioneering Research, Wako-shi, Saitama 351-0198, Japan}

\date{\today}

\begin{abstract}
The active harnessing of quantum resources in engineered quantum devices poses unprecedented requirements on device control. Besides the residual interaction with the environment, causing environment-induced decoherence, uncontrolled parameters in the system itself---disorder---remains as a substantial factor limiting the precision and thus the performance of devices. These perturbations may arise, for instance, due to imperfect sample production, stray fields, or finite accuracy of control electronics. Disorder-dressed quantum evolution means a unifying framework, based on quantum master equations, to analyze how these detrimental influences cause deviations from the desired system dynamics. This description may thus contribute to unveiling and mitigating disorder effects towards robust schemes. To demonstrate the broad scope of this framework, we evaluate two distinct scenarios: a central spin immersed in an isotropic spin bath, and a random mass Dirac particle. In the former scenario, we demonstrate how the disorder average reflects purity oscillations, indicating the time- and state-dependent severity of the disorder impact. In the latter scenario, we discuss disorder-induced backscattering and disorder-induced {\it Zitterbewegung} as consequences of the breakup of spin-momentum locking.
\end{abstract}

\maketitle

\section{Introduction}
The transformation of quantum science into an application-oriented engineering discipline comes with the promise of groundbreaking technologies, ranging from sensors with unprecedented precision, to spintronics, to communication and computing devices with quantum principles at their core. A diverse family of highly controllable systems, leveraging trapped ions \cite{Leibfried2003quantum, Blatt2012quantum, Britton2012engineered}, ultracold gases \cite{Bloch2012quantum, Wenz2013from, Jotzu2014experimental}, superconducting qubits \cite{Clarke2008superconducting, Leek2010cavity, Buluta2011natural, Boixo2014evidence, Kandala2017hardware}, quantum dots \cite{Hanson2007spins, Zwanenburg2013silicon, Lodahl2017quantum, Vandersypen2017interfacing}, spin impurities in solids \cite{Doherty2013nitrogen, Hanson2008coherent, Awschalom2018quantum}, photonics \cite{Raimond2001manipulating, Mair2001entanglement, OBrien2009photonic, Spring2013boson, Broome2013photonic, Crespi2013integrated, He2017time, Joannopoulos1997photonic, Ozawa2018topological}, and polaritons \cite{Carusotto2013quantum, Schneider2016exciton, Klembt2018exciton}, to name a few, is being developed to deliver the basic building blocks for the storage, processing, and transport of quantum states.

Achieving and upholding the desired functionality of these devices pose enormous challenges for system preparation, isolation, and control: Any accidental interaction with the environment, i.e., decoherence, can rapidly deteriorate the quantum resources, which is usually counteracted by cooling and isolating the systems. Similarly, uncontrolled variations of system parameters, disorder, while maintaining quantum coherence, can have a significant detrimental impact on the functionality of devices, in that they distort their intended functionality. These variations may be caused, e.g., by impurities in the sample, stray fields, or limitations in their external control; for instance, accidental gate overrotations in quantum computing devices. In many cases, such disorder constitutes one of the dominant remaining sources of error \cite{Hughes2005extrinsic, Wei2008controllable, Stark2011localization, Rogers2014multiple, Roux2015entanglement, Couto2011charge, Kuhlmann2013charge, Delbecq2016quantum, Reiner2018effects, Albash2018analog, Brugger2018quantum, Klimov2018fluctuations, Degen2017quantum, Ra2017reversed}.

The framework of disorder-dressed quantum evolution aims to capture and characterize the disorder-induced deviations of quantum systems from their intended dynamics. This is accomplished in terms of quantum master equations. The disorder impact on the evolution of the disorder-averaged state can then be understood, in analogy to the effect of a quantum environment, in terms of the---in general incoherent---deviations from the desired system dynamics. Understanding these deviations may then not only help to unveal fundamental disorder effects, but also contribute to the error analysis and mitigation in emerging quantum technologies \cite{Gneiting2017disorder, Gneiting2018lifetime, Gneiting2018disorder}. Mitigation of disorder-induced errors is, for instance, reflected in the design of transmon qubits (charge noise suppression) \cite{Schreier2008suppressing}, topological insulators (backscattering-immune edge transport) \cite{Hasan2010colloquium, Qi2011topological}, and variational-Hamiltonian hybrid algorithms (gate error mitigation) \cite{Colless2018computation}.

Disorder-dressed evolution is based on the disorder-averaged quantum state. On the one hand, this is motivated by the desire to identify statistically robust, generic disorder effects, the peculiarities of individual disorder realizations stripped off. On the other hand, this often corresponds to the situation realized in experiments, where disorder configurations, e.g., stray fields, fluctuate between runs. But even if the disorder is ``quenched'', disorder-dressed evolution allows one to capture the disorder effect, i.e., the deviation from the expected behavior, generically, independent of specific disorder realizations.

While individual disorder realizations describe coherent time evolution, i.e., pure states remain pure, ensemble averaging in general gives rise to varying state coherence \cite{Gneiting2016incoherent, Kropf2016effective}. The latter then indicates how different disorder realizations cause deviating state trajetories. In this sense, the coherence/purity of the averaged state carries information about the degree of the disorder-induced spread about the unperturbed trajectory, i.e., the variance among the perturbed trajectories. This feature, which has no correspondence in classical averaged states, then allows one to assess the disorder impact in terms of the purity of the averaged state. Ultimately, knowledge of the ensemble-averaged state $\overline{\rho}$ allows calculation of the disorder average of any observable $\hat{A}$, by virtue of $\int \dd \varepsilon \, p_\varepsilon \Tr [\rho_\varepsilon \hat{A}] = \Tr [\overline{\rho} \hat{A}]$, where $\rho_\varepsilon$ denotes the states for individual disorder realizations labeled by $\varepsilon$ and occurring with probability $p_\varepsilon$.

Our focus on static disorder, i.e., temporally unbounded correlations within individual disorder realizations, stands in contrast to the vanishing temporal correlations in the Markovian noise limit. Such lasting temporal correlations give rise to rich and possibly expedient non-Markovian dynamics (e.g., coherence/purity revivals \cite{Kropf2016effective} (in contrast to the strict purity decrease in the Markovian case \cite{Lidar2006conditions}) or bounded disorder-induced dephasing \cite{Gneiting2017disorder}), which has recently also come under intense scrutiny in the context of open quantum systems \cite{Chruscinski2010nonmarkovian, Liu2011experimental, Zhang2012general, LoFranco2012revival, Rivas2014quantum, Breuer2016colloquium, Addis2016problem}. Our approach aims at identifying dynamical effects associated with such temporal correlations, as well as with any other correlations within and among the disorder realizations.

A quantum master equation formulation for disorder dynamics was initially addressed in the limit of short times \cite{Gneiting2016incoherent}. Subsequently, it was shown that it can be determined (and solved) exactly for specific, symmetric disorder configurations \cite{Kropf2016effective}. This is however not the case for most generic scenarios, where the disorder interferes nontrivially with the system dynamics, while the short-time limit is too restrictive to capture many relevant disorder effects. On the other hand, the disorder contribution, which usually is deliberatively suppressed, can generically be considered to be small. We thus embrace a perturbative-in-the-disorder approach. While this excludes non-perturbative disorder effects, such as weak or strong localization at asymptotic times in transport scenarios, it comprises the disorder impact on the full quantum state, i.e., any (perturbative) disorder effect on observables is preserved and can be retrieved, e.g., the localization length encoded in the backscattering behavior, or the disorder-induced dephasing. From the perspective of quantum devices, with disorder effects inherently required to be small, restriction to the validity range of a perturbative approach appears justified, and a comprehensive description of the disorder impact, as delivered, desirable. In general, we can expect that the approximation remains valid beyond the point where the disorder impact exceeds acceptable thresholds. A more technical advantage arises from the fact that a perturbative expansion on the level of the evolution equation, as pursued here, produces, when solved, an improved approximation compared to an approximation on the level of the state/observable in a standard Born approximation.

The general form of the perturbative disorder-dressed evolution equation was introduced in \cite{Gneiting2017quantum}, where it was worked out with the example of a particle propagating in a disorder-perturbed wave guide, causing disorder-induced dephasing and backscattering. Subsequently, it has been applied to the edge-mode propagation in topological insulators, for a single \cite{Gneiting2017disorder} and two entangled \cite{Gneiting2018disorder} particles, and to a stability analysis of flatband states \cite{Gneiting2018lifetime}. In the present paper, the derivation of the general perturbative disorder-dressed evolution equation, based on the coupled-disorder-channel ansatz, is elaborated in detail. To further demonstrate its broad application range, we then evaluate it for two distinct scenarios: a central spin, immersed in a cloud of environmental spins (described by an isotropically randomized classical potential), and the random mass Dirac model, i.e., a massless Dirac particle, subject to spin-flipping perturbations. The former example characterizes several of the fundamental building blocks of quantum sensors or quantum computing devices, e.g., quantum dots or spin impurities in solids; the latter is a relevant model in many contexts of condensed matter, e.g., random spin chains, organic conductors, and quantum spin Hall edges. The latter example also serves to demonstrate how the emerging evolution equations can be solved efficiently with the help of the quantum phase space formalism.

Several highly sophisticated and successful theoretical tools exist to address disorder physics, including Green's function methods, transfer matrix implementations, and renormalization group approaches, some of them in particular excelling in the asymptotic-time and/or non-perturbative regime \cite{Lifshits1988introduction, Rammer1991quantum, Beenakker1997random}. Disorder-dressed evolution equations are meant to complement these, in the sense of capturing the onset of disorder effects comprehensively and in the time domain for arbitrary initial states, applicable to a wide range of disorder configurations and correlations, and tailored towards applications which build upon the preservation of quantum resources.

\section{Coupled disorder channels}

We begin by deriving the coupled disorder channel equations for general Hamiltonian ensembles. Disordered quantum systems can be characterized in terms of Hamiltonian ensembles, which characterize the lack of knowledge about and/or control of the system Hamiltonian. A general Hamiltonian ensemble $\{(\hat{H}_\varepsilon,p_\varepsilon)\}$ is comprised of a set of (in general arbitrary) Hamiltonians $\hat{H}_\varepsilon$, acting on the same quantum system and occurring with probability $p_\varepsilon$. The (multi-)index $\varepsilon$ may label a continuous, discrete, or finite set (or combinations thereof) of elements. Unless specified otherwise, we assume a continous probability distribution and write integrals, e.g., $\int \dd p_\varepsilon p_\varepsilon=1$. As a basic example, one may think of a single spin exposed to a magnetic field that varies slightly from run to run, $\{(\hat{H}_\varepsilon = (B_0+\varepsilon \Delta B) \sigma_z,p_\varepsilon)\}$, cf. \cite{Kropf2016effective, Chen2018simulating}. In the context of disordered quantum systems, it is useful to rewrite the Hamiltonians as $\hat{H}_{\varepsilon} = \hat{\overline{H}} + \hat{V}_{\varepsilon}$, where the averaged Hamiltonian $\hat{\overline{H}} \equiv \int \dd \varepsilon \, p_{\varepsilon} \, \hat{H}_{\varepsilon}$ describes the intended system behavior, and the disorder ``potentials'' $\hat{V}_\varepsilon$ (for convenience, we use this terminology in the general case), with $\int \dd \varepsilon \, p_{\varepsilon} \, \hat{V}_{\varepsilon} = 0$, capture uncontrolled perturbations, which cause deviations from the desired behavior. Single realizations are conceived as closed quantum systems following the von Neumann equation,
\begin{align} \label{Eq:von_Neumann_equation}
\partial_t \rho_{\varepsilon} = -\frac{\ii}{\hbar} [\hat{H}_{\varepsilon},\rho_{\varepsilon}] ,
\end{align}
which is formally solved for an arbitrary initial state $\rho_0$ (which we assume to be the same for all realizations) in terms of the time evolution operator $\hat{U}_{\varepsilon}(t) = \exp[-(\ii/\hbar) \hat{H}_{\varepsilon} t]$: $\rho_{\varepsilon}(t) = \hat{U}_{\varepsilon}(t) \rho_0 \hat{U}_{\varepsilon}^{\dagger}(t)$.

To analyze the disorder impact in a statistically robust way, devoid of nongeneric features present in single realizations, we consider the disorder-averaged state
\begin{align}
\overline{\rho}(t) \equiv \int \dd \varepsilon \, p_{\varepsilon} \, \rho_{\varepsilon}(t) = \int \dd \varepsilon \, p_{\varepsilon} \, \hat{U}_{\varepsilon}(t) \rho_0 \hat{U}_{\varepsilon}^{\dagger}(t) .
\end{align}
If we decompose $\rho_{\varepsilon} = \overline{\rho} + \Delta\rho_{\varepsilon}$ and take the ensemble average of the von Neumann equation (\ref{Eq:von_Neumann_equation}), we obtain the evolution equation
\begin{subequations} \label{Eq:coupled_disorder_channels}
\begin{align} \label{Eq:average_evolution}
\partial_t \overline{\rho}(t) = -\frac{\ii}{\hbar} [\hat{\overline{H}}, \overline{\rho}(t)] - \frac{\ii}{\hbar} \int \dd \varepsilon \, p_{\varepsilon} \, [\hat{V}_{\varepsilon}, \Delta\rho_{\varepsilon}(t)] .
\end{align}
We find that the dynamics of the averaged state $\overline{\rho}$ is not described by the averaged Hamiltonian $\hat{\overline{H}}$ alone, but modified by the coupling to the individual offsets $\Delta\rho_{\varepsilon}$, caused by the disorder potentials $\hat{V}_\varepsilon$. Indeed, the evolution of the disorder-averaged state in general transcends the unitary dynamics governing individual realizations.

The evolution equations for the offsets $\Delta\rho_{\varepsilon}$ are obtained by rewriting $\partial_t \Delta\rho_{\varepsilon} = \partial_t \rho_\varepsilon - \partial_t \overline{\rho}$ and applying (\ref{Eq:von_Neumann_equation}) and (\ref{Eq:average_evolution}):
\begin{align} \label{Eq:offset_evolution}
\partial_t \Delta\rho_{\varepsilon}(t) + \frac{\ii}{\hbar} [\hat{H}_{\varepsilon}, \Delta\rho_{\varepsilon}(t)] =& -\frac{\ii}{\hbar} [\hat{V}_{\varepsilon},\overline{\rho}(t)] \\
&+\frac{\ii}{\hbar} \int \dd \varepsilon' \, p_{\varepsilon'} \, [\hat{V}_{\varepsilon'}, \Delta\rho_{\varepsilon'}(t)] . \nonumber
\end{align}
\end{subequations}
The source terms on the right-hand-side describe the coupling to the averaged state and to the other offsets, respectively. Note that, in contrast to the realizations $\rho_\varepsilon$, the offsets $\Delta\rho_{\varepsilon}$ are dynamically coupled, which is a consequence of their common influence on the averaged state, and motivates the terminology of the ``coupled-disorder-channel equations'' (\ref{Eq:coupled_disorder_channels}). The corresponding initial conditions are $\overline{\rho}(t=0) = \rho_0$ and $\Delta\rho_{\varepsilon}(t=0) = 0, \, \forall \varepsilon$. Note that the offsets $\Delta\rho_{\varepsilon}$, in contrast to $\overline{\rho}$, do not describe normalized quantum states: $\Tr [\Delta\rho_{\varepsilon}] = 0$.

We remark that the coupled-disorder-channel equations (\ref{Eq:coupled_disorder_channels}), which are derived without any approximation, can be seen as a generalization of the Nakajima-Zwanzig projection operator technique \cite{Nakajima1959quantum, Zwanzig1960ensemble, Breuer2002theory}, with each disorder realization giving rise to an independent irrelevant component.

In the short-time limit, i.e., in the vicinity of $t=0$, where $\Delta\rho_{\varepsilon}(t) \approx 0$, (\ref{Eq:offset_evolution}) reduces to $\partial_t \Delta\rho_{\varepsilon}(t) \approx -\frac{\ii}{\hbar} [\hat{V}_{\varepsilon},\overline{\rho}(t)]$, which is solved by $\Delta\rho_{\varepsilon}(t) \approx -\frac{\ii}{\hbar} [\hat{V}_{\varepsilon},\overline{\rho}(t)] t$. Inserting this into (\ref{Eq:average_evolution}) recovers the short-time master equation derived, based on a different reasoning, in \cite{Gneiting2016incoherent}.

With the initial condition $\Delta\rho_{\varepsilon}(t=0) = 0$, the formal solution of (\ref{Eq:offset_evolution}) is determined, using the Green's formalism, by the inhomogeneous contribution alone, yielding
\begin{align} \label{Eq:formal_offset_solution}
\Delta\rho_{\varepsilon}(t) = \int_{0}^{t} \dd t' & U_\varepsilon(t-t') \Big\{-\frac{\ii}{\hbar} [\hat{V}_{\varepsilon},\overline{\rho}(t')] \\
&+ \frac{\ii}{\hbar} \int \dd \varepsilon' \, p_{\varepsilon'} \, [\hat{V}_{\varepsilon'}, \Delta\rho_{\varepsilon'}(t')] \Big\} U_\varepsilon^\dagger(t-t') . \nonumber
\end{align}
Iteratively inserting this solution into the second line of (\ref{Eq:formal_offset_solution}) gives rise to a Neumann series, which can be truncated at a desired order in the disorder potential $\hat{V}_{\varepsilon}$. If we insert the truncated solution into (\ref{Eq:average_evolution}), we then obtain a closed, perturbative, time-nonlocal evolution equation for the averaged state $\overline{\rho}$.

For some disorder configurations, the Neumann series (\ref{Eq:formal_offset_solution}) can be evaluated to infinite order, which then yields an exact evolution equation for the averaged state $\overline{\rho}$. This is, for instance, the case, if all disorder realizations $\hat{H}_\varepsilon$ commute, $[\hat{H}_\varepsilon, \hat{H}_{\varepsilon'}]=0 \; \forall \varepsilon,\varepsilon'$. This situation describes, e.g., an isolated flatband with potential disorder.

Let us remark that, in cases where the averaged state $\overline{\rho}(t) = \int \dd \varepsilon \, p_{\varepsilon} \, \hat{U}_{\varepsilon}(t) \rho_0 \hat{U}_{\varepsilon}^{\dagger}(t)$ can be evaluated directly, exact, time-local master equations can be derived by direct inversion of the corresponding dynamical map \cite{Andersson2007finding, Hall2014canonical, Kropf2016effective}. This was demonstrated, e.g., for the case of an ensemble of commuting Hamiltonians \cite{Kropf2016effective}.

\section{Disorder-perturbed dynamics}

Generically, the uncontrolled component of the Hamiltonian, i.e., the disorder, is weak compared to the target Hamiltonian, motivating a treatment perturbative in $\hat{V}_\varepsilon$. In order to obtain an evolution equation for $\overline{\rho}$ which is second order in the disorder potential $\hat{V}_\varepsilon$, we approximate (\ref{Eq:formal_offset_solution}) to first order in $\hat{V}_\varepsilon$. With $\overline{\rho}(t-\Delta t) = \hat{\overline{U}}(\Delta t)^\dagger \overline{\rho}(t) \hat{\overline{U}}(\Delta t) + \mathcal{O}(\hat{V}_{\varepsilon})$, this yields $\Delta\rho_{\varepsilon}(t) = -\frac{\ii}{\hbar} \int_{0}^{t} \dd t' \, [\hat{\tilde{V}}_{\varepsilon}(t'), \overline{\rho}(t)]$, where $\hat{\tilde{V}}_{\varepsilon}(t) = \hat{\overline{U}}(t) \hat{V}_{\varepsilon} \hat{\overline{U}}(t)^{\dagger}$ and $\hat{\overline{U}}(t) = \exp(- \ii \hat{\overline{H}} t/\hbar)$. Inserting this into (\ref{Eq:average_evolution}) then results in
\begin{align} \label{Eq:disorder_Redfield}
\partial_t \overline{\rho}(t) =& -\frac{\ii}{\hbar} [\hat{\overline{H}}, \overline{\rho}(t)] \nonumber \\
& - \frac{1}{\hbar^2} \int \dd \varepsilon \, p_{\varepsilon} \int_{0}^{t} \dd t' \, [\hat{V}_{\varepsilon}, [\hat{\tilde{V}}_{\varepsilon}(t'), \overline{\rho}(t)]] ,
\end{align}
which provides us with a closed dynamical equation for the disorder-averaged state $\overline{\rho}$. Note that this master equation is reminiscent of the Redfield equation, which captures the influence of a quantum environment on a quantum system after tracing out the environment \cite{Redfield1965theory, Breuer2002theory}. Here, the tracing operation is replaced by the disorder average. We stress that, despite this resemblance, we derived Eq.~(\ref{Eq:disorder_Redfield}) without reference to an actual or auxiliary environment, but by virtue of the coupled disorder channels (\ref{Eq:coupled_disorder_channels}). In contrast to the Redfield equation, which can, due to rapidly decaying bath correlations, often be simplified by taking the limit $t \rightarrow \infty$, this is not possible here. This reflects the non-Markovian nature of the disorder-averaged evolution, where individual disorder realizations display unconstrained temporal correlations. We remark that, due to its perturbative nature, the master equation (\ref{Eq:disorder_Redfield}) in general exhibits a finite temporal validity range. Moreover, in the limit $\hat{\tilde{V}}_\varepsilon(t) \approx \hat{V}_\varepsilon$, it reduces to the short-time master equation. Finally, we note that a related master equation can be obtained, under the assumption of classical stochastic, Gaussian noise, for noise-averaged states \cite{Chenu2017quantum}.

Using the identity
\begin{align}
[\hat{A},[\hat{B},\hat{X}]] = \frac{1}{2} [[\hat{A},\hat{B}],\hat{X}]
&- \frac{1}{4} [\hat{A}+\hat{B},[\hat{X},\hat{A}+\hat{B}]] \nonumber \\
&+ \frac{1}{4} [\hat{A}-\hat{B},[\hat{X},\hat{A}-\hat{B}]] ,
\end{align}
we can recast the master equation (\ref{Eq:disorder_Redfield}) in Lindblad structure; the latter reflects general quantum evolution beyond the von Neumann equation, consistent with the postulates of quantum mechanics. Moreover, this allows us to discuss coherent and incoherent contributions to the dynamics, to assess the positivity of the evolution, and, possibly, to interpret the dynamics in terms of the physical processes captured by the Lindblad operators, which may, e.g., be familiar from open systems. For instance, the Lindblad representation can render it manifest if symmetries that may be lost in single disorder realizations resurface in the collective behavior. We obtain
\begin{subequations} \label{Eq:perturbative_master_equation}
	\begin{align}
		\partial_t \overline{\rho}(t) =& -\frac{\ii}{\hbar} [\hat{H}_{\rm eff}(t), \overline{\rho}(t)] \nonumber \\
		& +\sum_{\alpha \in \{ \pm 1 \}} \frac{2 \alpha}{\hbar^2} \int \dd \varepsilon \, p_{\varepsilon} \int_{0}^{t} \dd t' \mathcal{L}\big(\hat{L}_{\varepsilon, t'}^{(\alpha)}, \overline{\rho}(t)\big) ,
	\end{align}
with $\mathcal{L}(\hat{L}, \rho) \equiv \hat{L} \rho \hat{L}^{\dagger} - \frac{1}{2} \hat{L}^{\dagger} \hat{L} \rho - \frac{1}{2} \rho \hat{L}^{\dagger} \hat{L}$. The corresponding (in general time-dependent) effective Hamiltonian $H_{\rm eff}(t) = H_{\rm eff}^{\dagger}(t)$ and Lindblad operators $\hat{L}_{\varepsilon, t}^{(\alpha)}$ read
	\begin{align}
		\hat{H}_{\rm eff}(t) &= \hat{\overline{H}} -\frac{\ii}{2 \hbar} \int \dd \varepsilon \, p_{\varepsilon} \int_{0}^{t} \dd t' \, [\hat{V}_{\varepsilon}, \hat{\tilde{V}}_{\varepsilon}(t')] , \nonumber \\
		\hat{L}_{\varepsilon, t}^{(\alpha)} &= \frac{1}{2} \big[\hat{V}_{\varepsilon} + \alpha \, \hat{\tilde{V}}_{\varepsilon}(t) \big] .
	\end{align}
\end{subequations}
Note that, according to this representation, each disorder realization gives rise to an independent decoherence term. Alternative, more compact, representations are often available by reexpressing the disorder integral in terms of the disorder correlation function; demonstrations of this appear below.

The $\alpha=-1$ contributions in (\ref{Eq:perturbative_master_equation}) describe negative decoherence ``rates'', indicating the feedback of coherence into the system, which, in turn, reflects the non-Markovian nature of the evolution. The corresponding Lindblad operators $\hat{L}_{\varepsilon, t}^{(-)}$ only build up slowly with time, $\hat{L}_{\varepsilon, t=0}^{(-)}=0$, consistent with the positivity of the evolution, and in agreement with the short-time master equation \cite{Gneiting2016incoherent}. Growth of the $\hat{L}_{\varepsilon, t}^{(-)}$, on the other hand, is required to reproduce the resurgence of, e.g., the state purity, a characteristic aspect of disorder-averaged quantum evolution.

It is instructive to determine the next-to-leading order short-time master equation. Approximating $\hat{\tilde{V}}_{\varepsilon}(t) = \hat{V}_\varepsilon + \frac{\ii}{\hbar} t [\hat{V}_\varepsilon,\hat{\overline{H}}] + \mathcal{O}(t^2)$, we obtain the simplified expression
\begin{align} \label{Eq:next-to-leading_short-time_ME}
\partial_t \overline{\rho}(t) = -\frac{\ii}{\hbar} [\hat{H}_{\rm eff}(t), \overline{\rho}(t)] + \frac{2 t}{\hbar^2} \int \dd \varepsilon \, p_{\varepsilon} \mathcal{L}\big(\hat{L}_{\varepsilon}(t), \overline{\rho}(t)\big) ,
\end{align}
where $\hat{H}_{\rm eff}(t) = \hat{\overline{H}} +\frac{t^2}{4 \hbar^2} \int \dd \varepsilon \, p_{\varepsilon} [\hat{V}_{\varepsilon}, [\hat{V}_{\varepsilon},\hat{\overline{H}}]]$ and $\hat{L}_{\varepsilon}(t) = \exp(- \ii \hat{\overline{H}} t/4 \hbar) \hat{V}_{\varepsilon} \exp(\ii \hat{\overline{H}} t/4 \hbar)$. We find that, while the Lindblad operators remain time-dependent, in this limit, no negative decoherence rates occur, rendering the positivity of the evolution manifest.

Evaluating (\ref{Eq:next-to-leading_short-time_ME}) for a particle in a parabolic band and a statistically homogeneous disorder potential, i.e., average Hamiltonian $\hat{\overline{H}}=\hat{p}^2/2 m$ and disorder correlations $\int \dd \varepsilon \, p_{\varepsilon} \hat{V}_\varepsilon(x) \hat{V}_\varepsilon(x') \equiv C(x-x') = \int \dd q \, \ee^{\ii q (x-x')/\hbar} G(q)$ (cf. \cite{Gneiting2017quantum}), we obtain the translation-covariant master equation [$G(-q)=G(q)$] $\partial_t \overline{\rho}(t) =$
\begin{align}
& -\frac{\ii}{\hbar} \Big[ \frac{\hat{p}^2}{2 m}, \overline{\rho}(t) \Big] \\
& + \int_{-\infty}^{\infty} \!\!\!\!\! \dd q \, \frac{2 G(q) t}{\hbar^2} \left\{ \ee^{\frac{\ii}{\hbar} q \hat{x}} \ee^{-\frac{\ii}{\hbar} \frac{q}{m} \frac{t}{4} \hat{p}} \overline{\rho}(t) \ee^{\frac{\ii}{\hbar} \frac{q}{m} \frac{t}{4} \hat{p}} \ee^{-\frac{\ii}{\hbar} q \hat{x}} - \overline{\rho}(t) \right\} . \nonumber
\end{align}
The occurring incoherent processes have a clear physical interpretation, relating to, and consistently complementing, the corresponding short-time master equation \cite{Gneiting2016incoherent}: The momentum kicks $\exp[\frac{\ii}{\hbar} q \hat{x}]$ displayed by the latter are here complemented by (growing with time) spatial displacements $\exp[-\frac{\ii}{\hbar} \frac{q}{m} \frac{t}{4} \hat{p}]$, reflecting the time evolution induced by preceding momentum kicks. The solution of the full disorder-dressed evolution (\ref{Eq:perturbative_master_equation}) for this case is discussed in \cite{Gneiting2017quantum}.

\section{Central spin}

\begin{figure}[htb]
	\includegraphics[width=0.95\columnwidth]{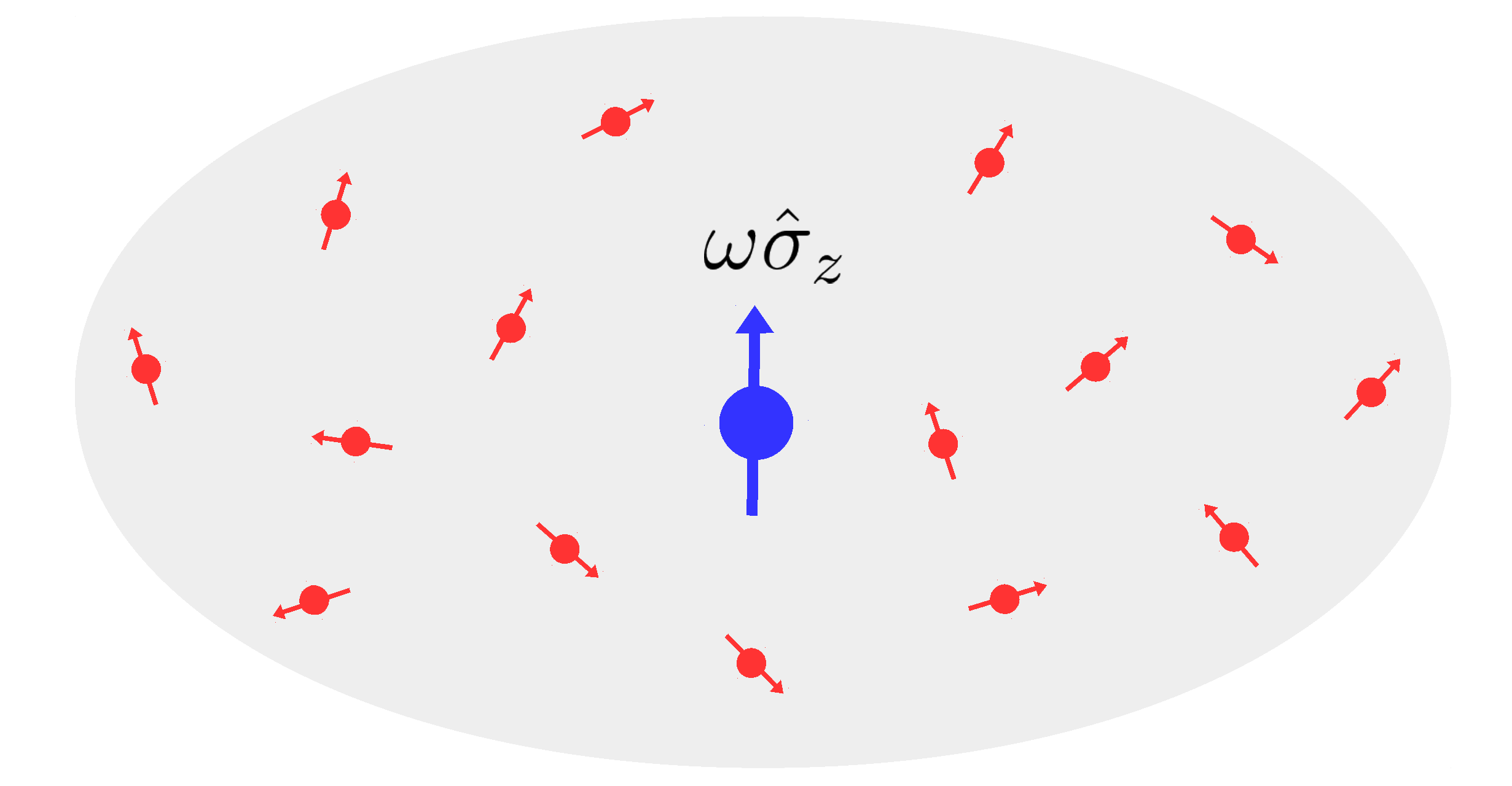}
	\caption{\label{Fig:Central_spin} Central spin immersed in a bath of classical, isotropically disordered spins. The central spin (blue arrow) is equipped with a control Hamiltonian $\hat{H}_0=\omega \hat{\sigma}_z$ aligned in the $z$ direction. A surrounding cloud of spins (red arrows) generates, while mostly averaging out, a residual, randomly oriented effective field, acting on the central spin as a disorder potential.}
\end{figure}

We now evaluate the disorder-dressed evolution equation (\ref{Eq:perturbative_master_equation}) for a central spin exposed to a classical, isotropically disordered environment, cf.~Fig.~\ref{Fig:Central_spin}. This may, e.g., describe the detrimental impact of randomly oriented environmental nuclear spins on solid-state qubits \cite{Khaetskii2002electron, Kuhlmann2013charge, Delbecq2016quantum, Yang2016quantum} (possibly in addition to applied noise mitigation strategies \cite{Puebla2013dynamic, Chekhovich2015suppression}), affecting the fidelity of quantum information processing protocols, or the deployment of these spins as quantum sensors \cite{Schoelkopf2003qubits, Degen2017quantum, Zhang2018improving}. A similar situation is treated in \cite{Kropf2016effective}, there however restricted to a degenerate (i.e., vanishing) system Hamiltonian, which may, e.g., correspond to an idling qubit, and which gives rise to isotropic depolarization dynamics. Here, we consider the more general case of a nondegenerate central spin equipped with a nonvanishing control Hamiltonian, lifting the isotropy of the environmental influence.

Without loss of generality, we assume that the spin Hamiltonian is aligned in the $z$ direction. Disordered Hamiltonians are then described by ($\hat{\sigma}_z = \ket{\uparrow}\bra{\uparrow} - \ket{\downarrow}\bra{\downarrow}$)
\begin{align} \label{Eq:central_spin_disordered_Hamiltonian}
\hat{H}_{W, \Delta} = \hbar \omega \hat{\sigma}_z + \frac{\Delta}{2} \hat{W} \hat{\sigma}_z \hat{W}^\dagger ,
\end{align}
where a single disorder configuration $\varepsilon$ is characterized by a random (unitary) rotation/orientation $\hat{W}$ of the environmental influence, drawn according to the Haar measure $\dd_\mu(W)$, along with a disordered level spacing $\Delta$, drawn from a probability distribution $p_\Delta$. It follows that $\hat{\overline{H}} = \hbar \omega \hat{\sigma}_z$ and $\hat{V}_{W,\Delta} = \frac{\Delta}{2} \hat{W} \hat{\sigma}_z \hat{W}^\dagger$, which then gives $\hat{\tilde{V}}_{W,\Delta}(t) = \frac{\Delta}{2} \ee^{-\ii \omega t \hat{\sigma}_z} \hat{W} \hat{\sigma}_z \hat{W}^\dagger \ee^{\ii \omega t \hat{\sigma}_z}$.

The corresponding master equation (\ref{Eq:perturbative_master_equation}) can be significantly simplified if we conduct the occurring Haar measure integrals, employing results from the Weingarten calculus for unitary groups \cite{Collins2006integration, Kropf2016effective}. Using
\begin{align}
\int \dd_\mu(W) & \hat{W} \hat{X}_1 \hat{W}^\dagger \hat{X}_2 \hat{W} \hat{X}_3 \hat{W}^\dagger \\
=& \frac{d \Tr[\hat{X}_1 \hat{X}_3] - \Tr[\hat{X}_1] \Tr[\hat{X}_3]}{d (d^2-1)} \Tr[\hat{X_2}] \mathbb{1} \nonumber \\
&+ \frac{d \Tr[\hat{X}_1] \Tr[\hat{X}_3] - \Tr[\hat{X}_1 \hat{X}_3]}{d (d^2-1)} \hat{X}_2 \nonumber
\end{align}
with $d=2$, $\hat{X}_1 = \hat{X}_3 = \hat{\sigma}_z$, and $\hat{X}_2 = \ee^{-\ii \omega t \hat{\sigma}_z}$, we evaluate the effective Hamiltonian $\hat{H}_{\rm eff}(t) = \hat{\overline{H}} -\frac{\ii}{2 \hbar} \int \dd_\mu(W) \int \dd\Delta \, p_\Delta \int_{0}^{t} \dd t' \, [\hat{V}_{W,\Delta}, \hat{\tilde{V}}_{W,\Delta}(t')]$ as
\begin{align} \label{Eq:central_spin_effective_Hamiltonian}
\hat{H}_{\rm eff}(t) = \hbar \omega \hat{\sigma}_z \left( 1-\frac{\overline{\Delta^2} t^2}{12 \hbar^2} \sinc^2[\omega t] \right) ,
\end{align}
where $\overline{\Delta^2} \equiv \int \dd\Delta \, p_\Delta \Delta^2$ (assuming that the distribution $p_\Delta$ exhibits a well-defined variance). We thus find that the disorder average induces a periodic modulation of the angular velocity of the spin rotation about the $z$ axis. Similarly, we obtain for the incoherent part of (\ref{Eq:perturbative_master_equation})
\begin{align}
& \sum_{\alpha \in \{ \pm 1 \}} \alpha \int \dd_\mu(W) \, \int \dd\Delta \, p_\Delta \mathcal{L}\big(\hat{L}_{W,\Delta,t'}^{(\alpha)}, \overline{\rho}(t)\big) = \\
& \frac{\overline{\Delta^2}}{6} \{\cos^2(\omega t') [\mathbb{1}_2 - 2 \overline{\rho}(t)] + \sin^2(\omega t') \Tr[\overline{\rho}(t) \hat{\sigma}_z] \hat{\sigma}_z \} . \nonumber
\end{align}
Note how here, as in the effective Hamiltonian (\ref{Eq:central_spin_effective_Hamiltonian}), the $z$ axis persists as a symmetry axis of the dynamics.

The compactified master equation can again be recast in Lindblad form, using $\Tr[\rho \hat{\sigma}_z] \hat{\sigma}_z = \mathcal{L}(\hat{P}_\uparrow,\rho) - \mathcal{L}(\hat{\sigma}_+,\rho) + \mathcal{L}(\hat{P}_\downarrow,\rho) - \mathcal{L}(\hat{\sigma}_-,\rho)$ and $\mathbb{1}_2-2 \rho = \mathcal{L}(\hat{P}_\uparrow,\rho) + \mathcal{L}(\hat{P}_\downarrow,\rho) + \mathcal{L}(\hat{\sigma}_+,\rho) + \mathcal{L}(\hat{\sigma}_-,\rho)$, which yields the disorder-dressed evolution equation
\begin{align} \label{Eq:central_spin_master_equation}
\partial_t \overline{\rho}(t) =& -\frac{\ii}{\hbar} [\hat{H}_{\rm eff}(t), \overline{\rho}(t)] \nonumber \\
&+ \frac{\overline{\Delta^2} t}{3 \hbar^2} \Big\{\mathcal{L}(\hat{P}_\uparrow,\overline{\rho}(t)) + \mathcal{L}(\hat{P}_\downarrow,\overline{\rho}(t)) \nonumber \\
&+ \sinc(2 \omega t) [\mathcal{L}(\hat{\sigma}_+,\overline{\rho}(t)) + \mathcal{L}(\hat{\sigma}_-,\overline{\rho}(t))] \Big\} .
\end{align}
The Lindblad operators are given by the level projectors $\hat{P}_\uparrow = \ket{\uparrow}\bra{\uparrow}$ and $\hat{P}_\downarrow = \ket{\downarrow}\bra{\downarrow}$, and the ladder operators $\hat{\sigma}_+ = \ket{\uparrow}\bra{\downarrow}$ and $\hat{\sigma}_- = \ket{\downarrow}\bra{\uparrow}$, and $\hat{H}_{\rm eff}(t)$ as in (\ref{Eq:central_spin_effective_Hamiltonian}). We thus find that the nonvanishing system Hamiltonian lifts the isotropy in the incoherent part of the dynamics, too. On the other hand, the rotational symmetry of the combination of system Hamiltonian and isotropic disorder is restored. In the limit $\omega \rightarrow 0$, we recover the isotropic depolarization dynamics induced by the, then remaining, isotropically disordered environment, $\partial_t \overline{\rho}(t) = \frac{2 \overline{\Delta^2} t}{3 \hbar^2} (\frac{1}{2} \mathbb{1}_2 - \overline{\rho}(t))$, which corresponds to the short-time limit of the exact evolution equation discussed in \cite{Kropf2016effective}.

The (non-Markovian) master equation (\ref{Eq:central_spin_master_equation}) can be solved exactly. We obtain for the diagonal and off-diagonal matrix elements ($\overline{\rho}_{\uparrow \uparrow} \equiv \bra{\uparrow} \rho \ket{\uparrow}$)
\begin{subequations} \label{Eq:central_spin_solution}
\begin{align}
\overline{\rho}_{\uparrow \uparrow}(t) = \frac{1}{2}+(\overline{\rho}_{\uparrow \uparrow,0}-1/2) \, \exp \left( -\frac{\overline{\Delta^2} t^2}{3 \hbar^2} \sinc^2[\omega t] \right) ,
\end{align}
and ($\overline{\rho}_{\uparrow \downarrow} \equiv \bra{\uparrow} \rho \ket{\downarrow}$)
\begin{align}
\overline{\rho}_{\uparrow \downarrow}(t) = \overline{\rho}_{\uparrow \downarrow,0} \,& \ee^{-2 \ii \omega t} \exp \left( \ii \frac{\overline{\Delta^2} t}{12 \hbar^2 \omega} (1-\sinc[2 \omega t]) \right) \nonumber \\
& \times \exp \left( -\frac{\overline{\Delta^2} t^2}{6 \hbar^2} (1+\sinc^2[\omega t]) \right) ,
\end{align}
\end{subequations}
respectively. We thus find that, within the limits of our approximation, the diagonal elements display ongoing oscillations, modulated by the spin frequency $\omega$, while the offdiagonal elements describe exponentially decaying Rabi oscillations, again modulated by oscillating correction terms.

\begin{figure}[htb]
	\includegraphics[width=0.90\columnwidth]{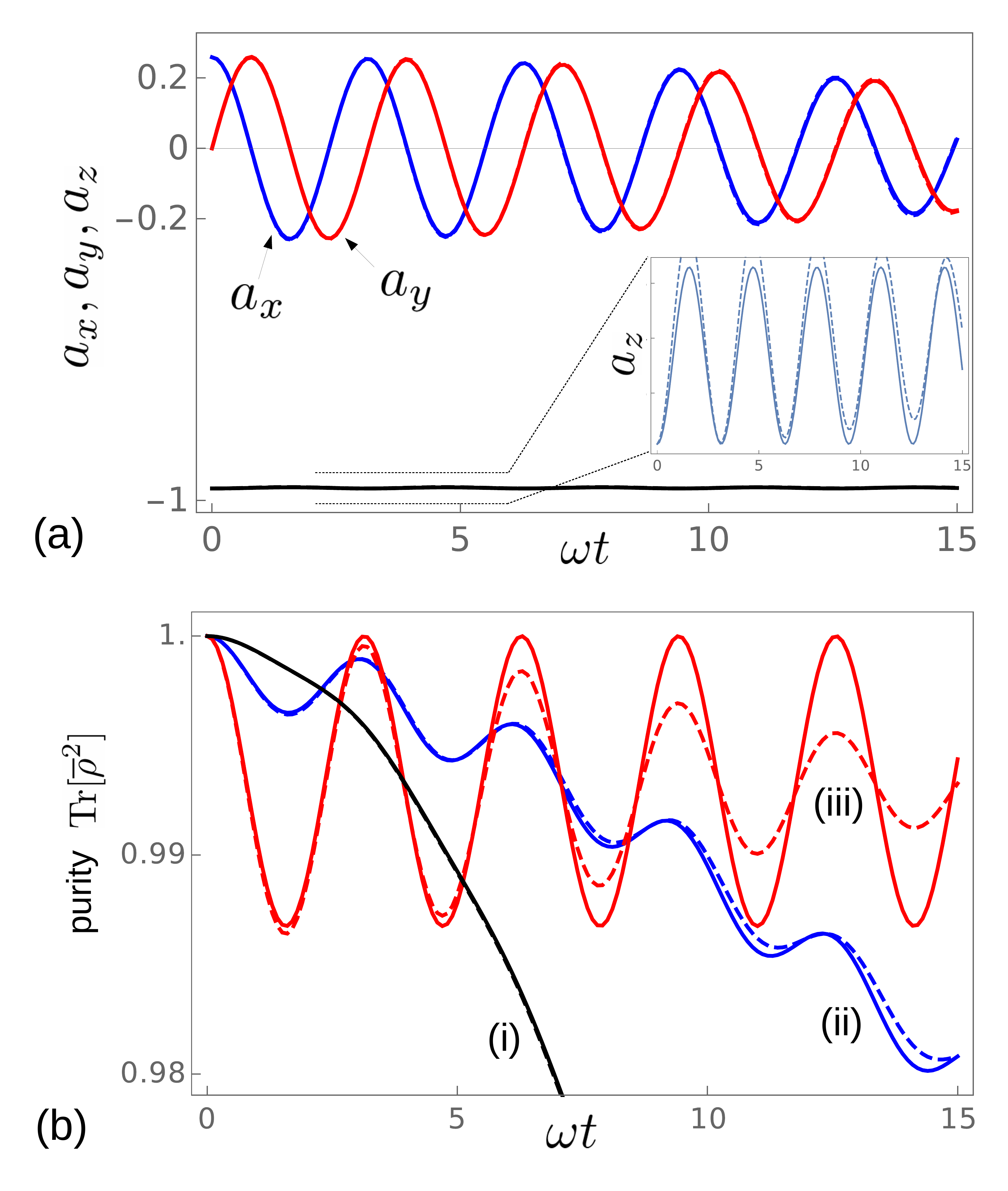}
	\caption{\label{Fig:Central_spin_dynamics} Disorder-dressed evolution of a central spin immersed in a bath of classical, isotropically disordered spins. (a) A generic initial state $\ket{\psi_0}=\cos(\pi/12) \ket{\downarrow}+ \sin(\pi/12) \ket{\uparrow}$ [case (ii) in (b)] displays temporally modulated Rabi oscillations in the $x$-$y$ plane of the Bloch sphere (Bloch vector $\vec{a}$), complemented by additional oscillations in the $z$ component (inset). The latter are of purely incoherent nature and arise as a consequence of the interplay between the disorder and the control Hamiltonian. Shown are the predictions of the disorder-dressed evolution equation (\ref{Eq:central_spin_master_equation}) [solid lines] and the directly ensemble-averaged evolution ($K=1000$ realizations) [dashed lines]. (b) The purity evolution, which reflects the amount of disorder-induced mixing, displays qualitatively and quantitatively different behavior for different initial states: Initial states (i) at the equator of the Bloch sphere display a strong, overall exponential decay of purity, with a temporally modulated decay rate. Initial states (iii) at the poles of the Bloch sphere exhibit (comparatively) weak, purely disorder-induced oscillations towards the center/maximally mixed state. (ii) Intermediate initial states display weighted combinations of these behavorial traits.}
\end{figure}

Figure~\ref{Fig:Central_spin_dynamics} shows, in terms of the Bloch vector $\vec{a}$, $\rho=(\mathbb{1}_2+\vec{a} \cdot \hat{\vec{\sigma}})/\sqrt{2}$, the time evolution for the three cases (i) $\sqrt{\overline{\Delta^2}}=0.05 \omega$ and $\ket{\psi_0}=(\ket{\downarrow}+\ket{\uparrow})/\sqrt{2}$ (initial state on the equator of the Bloch sphere), (ii) $\sqrt{\overline{\Delta^2}}=0.1 \omega$ and $\ket{\psi_0}=\cos(\pi/12) \ket{\downarrow}+ \sin(\pi/12) \ket{\uparrow}$ (initial state near the south pole of the Bloch sphere), and (iii) $\sqrt{\overline{\Delta^2}}=0.2 \omega$ and $\ket{\psi_0}= \ket{\downarrow}$ (initial state at the south pole of the Bloch sphere). We compare the prediction of the disorder-dressed evolution equation (\ref{Eq:central_spin_master_equation}) or (\ref{Eq:central_spin_solution}), respectively (solid lines), with the direct, numerically exact, ensemble-averaged evolution (dashed lines), averaged over $K=1000$ realizations of the disordered Hamiltonian (\ref{Eq:central_spin_disordered_Hamiltonian}) (with the realizations of the disorder potential $\hat{V}_{W, \Delta} = \frac{\Delta}{2} \hat{W} \hat{\sigma}_z \hat{W}^\dagger$ sampled from a Gaussian unitary ensemble). Shown are the time evolution of the Bloch vector components in case (ii), and the purity evolution for all three cases. The purity $r(t) = \Tr[\overline{\rho}(t)^2]$ serves as a useful quantifier for the disorder impact, measuring the averaging-induced state mixing \cite{Gneiting2016incoherent}. Purity revivals (full or partial), on the other hand, indicate the convergence of different disorder realizations in state space and can thus serve to identify and exploit disorder robustness.

We find that the disorder-dressed evolution equation describes the dynamics well in the short to intermediate time domain. All disorder-induced dynamical features are recovered by the direct averaging: In case (i), the Rabi oscillating state displays a strong, overall exponential decay of coherences, with a temporally modulated decoherence rate. In case (ii), the modulated Rabi oscillations are complemented by an oscillation of the $z$ component of the Bloch vector. The latter, which is of purely incoherent nature, is disorder-induced and arises as a consequence of the interplay between the control Hamiltonian and the disorder potential. If the initial state is located at one of the poles (which are fixed points of the disorder-free evolution), case (iii), these state-dependent incoherent oscillations remain as the sole dynamical trait. In this case, the purity coincides with the $z$ component of the Bloch vector $a_z(t)$. These oscillations are present neither in the absence of disorder nor in the absence of the control Hamiltonian. Note that, in case (iii), the numerically exact ensemble-averaged evolution exhibits damped purity oscillations, while this damping is not reflected by the evolution equation (\ref{Eq:central_spin_master_equation}). This is a consequence of the perturbative nature of (\ref{Eq:central_spin_master_equation}), where higher-order contributions of the disorder potential are neglected (for demonstrational purposes, we choose comparatively strong disorder potentials). This also limits the temporal validity of the described evolution.

As emphasized above, such analysis of the purity evolution of the disorder-averaged state, reflected here by state-dependent purity oscillations, may help, e.g., to identify optimal readout times in quantum sensing or gate applications (assuming noise to be static over the sensing or gate duration), contributing to minimizing the disorder impact. In the present case, these readout times would be chosen at local purity maxima. We remark that, while the displayed disorder-induced purity losses on the order of (for strong disorder) a few percent may appear small, present-day quantum devices are often concerned with fidelity/purity control in the deep subpercent regime.

\section{Massless Dirac particle}

\begin{figure}[htb]
	\includegraphics[width=0.99\columnwidth]{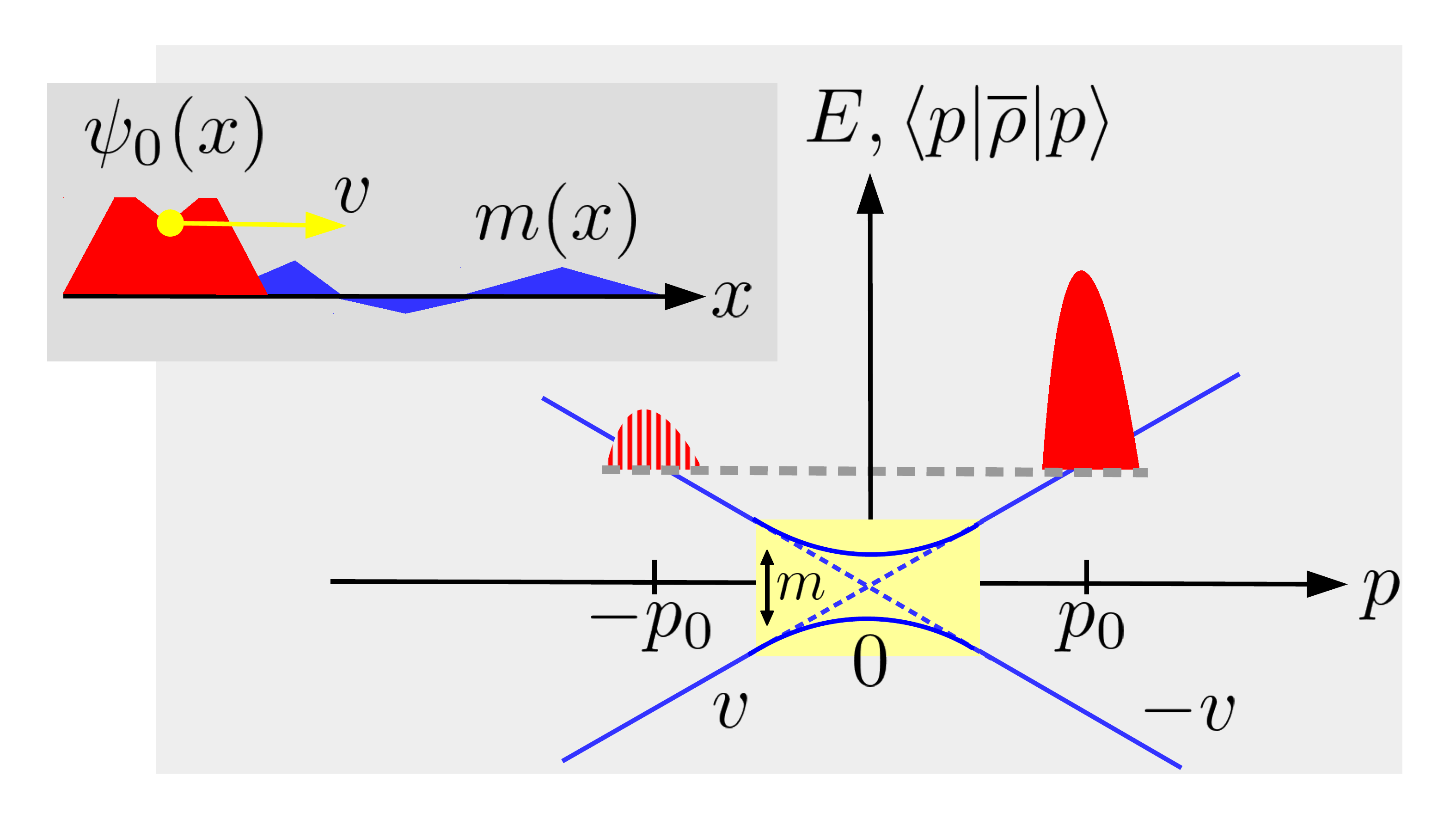}
	\caption{\label{Fig:Dirac_backscattering} Backscattering of relativistic Dirac/Weyl particles in the presence of disordered spin-flipping potentials. Small frame: A rightmoving initial state $\psi_0(x)$ (red shape) propagates along a spin-flipping potential (blue shape), formally captured by a spatially disordered mass term. Large frame: If the initial state is centered around $p_0$ in momentum space (red solid shape), then a fluctuating mass term (yellow area) gives rise to backscattering into the left moving band branch at $-p_0$ (red dashed shape). A scalar disorder potential, in contrast, would cause no backscattering.}
\end{figure}

As the second example, we now discuss a massless Dirac particle, confined to one dimension, and subject to a disordered mass term, see Fig.~\ref{Fig:Dirac_backscattering}. Besides its fundamental interest \cite{Thaller2013dirac, Lamata2007dirac, Cabrera2016dirac}, this random mass Dirac model approximates generic situations in condensed matter physics and spintronics, e.g., random spin chains or organic conductors \cite{Takayama1980continuum, Fisher1994random, Steiner1998random}, or helical edge states of topological insulators \cite{Hasan2010colloquium, Qi2011topological}. Apart from its natural occurrence in condensed matter systems, emulations of the random mass Dirac model are also available with, e.g., integrated optics \cite{Keil2013random} or ultracold atoms \cite{Edmonds2012anderson}. If there is on-site/potential disorder only, i.e., in the absence of mass perturbations, propagation is backscattering-free \cite{Katsnelson2006chiral, Beenakker2008colloquium}, and disorder-induced dephasing remains as a disorder effect, as discussed, e.g., in \cite{Gneiting2017disorder}. In contrast, as we derive now, perturbations in the mass term can give rise to backscattering (see also \cite{Hsu2017nuclear, Vaeyrynen2018noise} and references therein) and Zitterbewegung.

The starting point of our analysis is the one-dimensional Dirac Hamiltonian with mass perturbations (in the case of lattice systems we assume the continuum limit):
\begin{align}
\hat{H}_\varepsilon = v \, \hat{p} \, \hat{\sigma}_z + m_\varepsilon(\hat{x}) \, v^2 \, \hat{\sigma}_x ,
\end{align}
with drift velocity $v$, $\hat{\sigma}_z = \ket{\uparrow}\bra{\uparrow} - \ket{\downarrow}\bra{\downarrow}$, and $\hat{\sigma}_x = \ket{\uparrow}\bra{\downarrow} + \ket{\downarrow}\bra{\uparrow}$. In the case of helical edge electrons in topological insulators, one may think of the mass perturbations, e.g., as (pseudo-)spin-flipping magnetic impurities. If we assume on average vanishing mass fluctuations, $\int \dd \varepsilon \, p_{\varepsilon} \, m_\varepsilon(\hat{x}) = 0$, the average Hamiltonian reads $\hat{\overline{H}} = v \, \hat{p} \, \hat{\sigma}_z$. We further assume translation-invariant disorder correlations,
\begin{align} \label{Eq:translation-invariant_correlations}
C(x-x') \equiv & \int \dd \varepsilon \, p_{\varepsilon} m_\varepsilon(x) v^2 m_\varepsilon(x') v^2 \nonumber \\
= & \int \dd q \, \ee^{\ii q (x-x')/\hbar} G(q) ,
\end{align}
such that the disorder impact is summarized by the momentum transfer distribution $G(q)=G(-q)$. With $\hat{V}_\varepsilon = m_\varepsilon(\hat{x}) v^2 \otimes \sigma_x = \int \dd x \, m_\varepsilon(x) v^2 \ket{x}\bra{x} \otimes \sigma_x$ and using (\ref{Eq:translation-invariant_correlations}), we can rewrite (\ref{Eq:disorder_Redfield}) as
\begin{align} \label{Eq:momentum_Redfield}
\partial_t \overline{\rho}(t) =& - \frac{\ii}{\hbar} [\hat{\overline{H}}, \overline{\rho}(t)] \nonumber \\
&- \frac{1}{\hbar^2} \int \dd q \, G(q) \int_{0}^{t} \dd t' \, [\hat{V}_q, [\hat{\tilde{V}}^\dagger_{q}(t'), \overline{\rho}(t)]] ,
\end{align}
where $\hat{V}_q = \int \dd x \, \ee^{\frac{\ii}{\hbar} q x} \ket{x}\bra{x} \otimes \hat{\sigma}_x = \ee^{\frac{\ii}{\hbar} q \hat{x}} \otimes \hat{\sigma}_x$ (describing momentum kicks accompanied by simultaneous band swapping) and $\hat{\tilde{V}}_q(t) = \ee^{-\frac{\ii}{\hbar} \hat{\overline{H}} t} \hat{V}_q \ee^{\frac{\ii}{\hbar} \hat{\overline{H}} t}$. We remark that (\ref{Eq:momentum_Redfield}) can be brought into Lindblad form similarly to (\ref{Eq:perturbative_master_equation}), cf.~\cite{Gneiting2017disorder, Gneiting2018lifetime, Gneiting2018disorder}. For evaluations, it is often convenient to work with (\ref{Eq:momentum_Redfield}). In either case, the equation reflects the translational invariance that is restored on the collective level of the disorder average.

The time evolution operator can be rewritten as $\ee^{-\frac{\ii}{\hbar} \hat{\overline{H}} t} = \ee^{\frac{\ii}{\hbar} v t \hat{p}} \otimes \hat{P}_\downarrow + \ee^{-\frac{\ii}{\hbar} v t \hat{p}} \otimes \hat{P}_\uparrow$, which yields
\begin{align}
\hat{\tilde{V}}_q(t) = \ee^{-\frac{\ii}{\hbar} v t q} \big( \ee^{\frac{\ii}{\hbar} q \hat{x}} \ee^{2 \frac{\ii}{\hbar} v t \hat{p}} \otimes \hat{\sigma}_- + \ee^{-2 \frac{\ii}{\hbar} v t \hat{p}} \ee^{\frac{\ii}{\hbar} q \hat{x}} \otimes \hat{\sigma}_+ \big) ,
\end{align}
with $\hat{P}_\uparrow$, $\hat{P}_\downarrow$, $\hat{\sigma}_+$, and $\hat{\sigma}_-$ as in the previous section. For the band projection $\overline{\rho}_{\uparrow \uparrow} \equiv \bra{\uparrow} \rho \ket{\uparrow}$, we then obtain the evolution equation
\begin{align} \label{Eq:Dirac_upspin_evolution_equation}
\partial_t \overline{\rho}_{\uparrow \uparrow}(t) =& - \frac{\ii}{\hbar} [v \hat{p}, \overline{\rho}_{\uparrow \uparrow}(t)] \\
& -\int \dd q \, \frac{G(q)}{\hbar^2} \int_{0}^{t} \dd t' \big\{ \ee^{\frac{\ii}{\hbar} v t' q} \ee^{2 \frac{\ii}{\hbar} v t' \hat{p}} \overline{\rho}_{\uparrow \uparrow}(t) \nonumber \\
& -\ee^{\frac{\ii}{\hbar} v t' q} \ee^{\frac{\ii}{\hbar} q \hat{x}} \overline{\rho}_{\downarrow \downarrow}(t) \ee^{-\frac{\ii}{\hbar} q \hat{x}} \ee^{2 \frac{\ii}{\hbar} v t' \hat{p}} + h.c. \big\} . \nonumber
\end{align}
The corresponding equation for the opposite band component $\overline{\rho}_{\downarrow \downarrow} \equiv \bra{\downarrow} \rho \ket{\downarrow}$ takes the same form, with $v$ replaced by $-v$.

To proceed towards a solution of the master equation, it is useful to transform the coupled evolution equations for the two bands into phase space language. Indeed, it turns out that the phase space formalism, while often unfavorable for dynamical treatments, allows for comparatively simple and elegant solutions with the Lindblad terms arising in translationally-invariant disordered quantum systems, cf.~Eq.~(\ref{Eq:Dirac_upspin_evolution_equation}).

We briefly recapitulate the phase-space representation, which provides us with a self-consistent reformulation of quantum mechanics, equivalent to the standard operator-based formalism \cite{Wigner1932a, Moyal1949a, Weyl1928a, Schleich2011quantum, Gneiting2013quantum}. The transformation from operators to phase-space functions is accomplished with the help of the Stratonovich-Weyl operator kernel \cite{Stratonovich1957distributions}, which is defined as
\begin{align} \label{Eq:Stratonovich-Weyl_kernel}
\hat{\Delta}(x,p) = \hat{D}(x,p) \hat{\Delta}(0,0) \hat{D}^\dagger(x,p) ,
\end{align}
with the displacement operators
\begin{align}
\hat{D}(x,p) = \exp \left( -\frac{\ii}{\hbar} x \hat{p} \right) \exp \left( \frac{\ii}{\hbar} p \hat{x} \right) ,
\end{align}
and the undisplaced operator kernel
\begin{align}
\hat{\Delta}(0,0) = \int \dd x' \, \ket{x'/2} \bra{-x'/2} .
\end{align}
The latter is related to the parity operator $\hat{P}=\int \dd x \, \ket{x} \bra{-x}$, $\hat{\Delta}(0,0) = 2 \hat{P}$. This is why the Stratonovich-Weyl operator kernel is sometimes referred to as displaced parity.

Based on the kernel (\ref{Eq:Stratonovich-Weyl_kernel}), the Weyl symbol (i.e., phase space representation) $W_{\hat{A}}(x,p)$ of a general operator $\hat{A}$ is obtained according to
\begin{align} \label{Eq:Weyl_symbol}
W_{\hat{A}}(x,p) &= \Tr[\hat{A} \hat{\Delta}(x,p)] \\
&= \int\dd x' \, \ee^{\frac{\ii}{\hbar} p x'} \bra{x-x'/2} \hat{A} \ket{x+x'/2} . \nonumber
\end{align}
For the sake of normalization, a rescaled Weyl symbol, the Wigner function $W(x,p)$, is introduced for the density operator $\rho$, $W(x,p) = \frac{1}{2 \pi \hbar} W_{\rho}(x,p)$, which then satisfies $\int \dd x \dd p \, W(x,p) = 1$. In addition, the marginals of the Wigner function evaluate as $\int \dd p \, W(x,p) = \bra{x} \rho \ket{x}$ and $\int \dd x \, W(x,p) = \bra{p} \rho \ket{p}$, which motivates its interpretation as a quasi-probability distribution. However, the Wigner function can take negative values, which can be considered as a signature for quantumness.

Using (\ref{Eq:Weyl_symbol}), we can reexpress the evolution equation (\ref{Eq:Dirac_upspin_evolution_equation}) (and its opposite-band counterpart) in terms of the Wigner function:
\begin{align} \label{Eq.Dirac_Wigner_evolution_equation}
(\partial_t \pm v \, \partial_x) & W_t^{\pm}(x,p) = \nonumber \\
-\int \dd q'& \, \frac{2 G(q')}{\hbar^2} \int_0^t \dd t' \cos \left[ \frac{v t' (q'+2 p)}{\hbar} \right] \\
&\times \Big\{ W_t^{\pm}(x \pm v t', p) - W_t^{\mp}(x \mp v t', p-q') \Big\} , \nonumber
\end{align}
where $W_t^{+}(x,p)$ [$W_t^{-}(x,p)$] denotes the Wigner function of the right-(left-)moving state component $\overline{\rho}_{\uparrow \uparrow}$ ($\overline{\rho}_{\downarrow \downarrow}$), and now $\int \dd x\dd p \, (W_t^{+}(x,p) + W_t^{-}(x,p)) = 1$. Here, we exploit that the spatial and momentum translation operators in (\ref{Eq:Dirac_upspin_evolution_equation}) can be rearranged towards shifting the Stratonovich-Weyl operator kernel, with the help of the identity
\begin{align}
\ee^{-\ii \Delta x \hat{p}/\hbar} \hat{\Delta}(x,p) = \hat{\Delta}(x,p) \ee^{\ii \Delta x \hat{p}/\hbar} \ee^{-2 \ii \Delta x p/\hbar} .
\end{align}

To turn (\ref{Eq.Dirac_Wigner_evolution_equation}) into a local differential equation, we further transform the Wigner function into its characteristic function, $\chi(s,q) = \int \dd x \dd p \, \ee^{-\frac{\ii}{\hbar} (q x-p s)} W(x,p)$. Moreover, we assume that the initial state is centered around a momentum $p_0$ (without loss of generality $p_0>0$), in line with a large wavepacket approximation. This implies that the Wigner function, too, is focused around $p_0$, such that we can approximate $p \approx p_0$ in the cosine in (\ref{Eq.Dirac_Wigner_evolution_equation}). With this, we obtain the coupled evolution equations
\begin{align}
[\partial_t \pm \frac{\ii}{\hbar} v q] & \chi_t^\pm(s,q) = \nonumber \\
\int \dd q' & \frac{2 G(q')}{\hbar^2} \int_{0}^{t} \dd t' \cos \left[ \frac{v t' (q'+2 p_0)}{\hbar} \right] \\
& \times \left\{ \ee^{\mp \frac{\ii}{\hbar} q v t'} \ee^{\frac{\ii}{\hbar} q' s} \chi_t^{\mp}(s,q) - \ee^{\pm \frac{\ii}{\hbar} q v t'} \chi_t^\pm(s,q) \right\} . \nonumber
\end{align}
Rewriting these coupled equations in terms of a single matrix equation,
\begin{align}
[\mathbb{1}_2 \partial_t +& \frac{\ii}{\hbar} v q \sigma_z] \vec{\chi}_t(s,q) = \nonumber \\
& \left( \begin{array}{cc} -F_t(0,-q) & F_t(s,q) \\ F_t(s,-q) & -F_t(0,q) \end{array} \right) \vec{\chi}_t(s,q) ,
\end{align}
where $\vec{\chi}_t(s,q) = (\chi_t^+(s,q),\chi_t^-(s,q))^T$ and
\begin{align}
& F_t(s,q) = \nonumber \\
& \int \dd q' \frac{2 G(q')}{\hbar^2} \int_0^t \dd t' \cos \left[ \frac{v t' (q'+2 p_0)}{\hbar} \right] \ee^{-\frac{\ii}{\hbar} q v t'} \ee^{\frac{\ii}{\hbar} q' s} ,
\end{align}
the resulting solution reads
\begin{align} \label{Eq:Dirac_characteristic_spinor_solution}
\vec{\chi}_t(s,q) =& \nonumber \\
\exp \Big[&-\overline{F}_t^{(g)}(0,q) \mathbb{1}_2 + \overline{F}_t^{(g)}(s,q) \sigma_x - \overline{F}_t^{(u)}(s,q) \sigma_y \nonumber \\
&-\ii \Big( \frac{v t q}{\hbar} - \overline{F}_t^{(u)}(0,q) \Big) \sigma_z \Big] \vec{\chi}_0(s,q) .
\end{align}
Here, we have decomposed the disorder influence $\overline{F}_t(s,q) \equiv \int_0^t \dd t' F_{t'}(s,q)$ into an even function $\overline{F}_t^{(g)}(s,-q) = \overline{F}_t^{(g)}(s,q)$ and an odd function $\overline{F}_t^{(u)}(s,-q) = -\overline{F}_t^{(u)}(s,q)$, $\overline{F}_t(s,q) = \overline{F}_t^{(g)}(s,q) + \ii \overline{F}_t^{(u)}(s,q)$. In particular, one then obtains
\begin{align} \label{Eq:Dirac_disorder_impact_even_component}
\overline{F}_t^{(g)}(s,q) = \int \dd q' \frac{t^2 G(q')}{2 \hbar^2} & \ee^{\frac{\ii}{\hbar} q' s} \bigg\{ \sinc^2 \left[ \frac{v t (2 p_0+q+q')}{2 \hbar} \right] \nonumber \\
+ \sinc^2 & \left[ \frac{v t (2 p_0-q+q')}{2 \hbar} \right] \bigg\} .
\end{align}
If we further assume a finite correlation length $\ell$ in the disordered mass fluctuations, and a finite position uncertainty $\sigma$ of the initial state, we can, in the time limit $v t \gg \ell, \sigma$, approximate
\begin{align} \label{Eq:Dirac_disorder_impact_large_time}
\overline{F}_t^{(g)}(s,q) = \frac{\pi t}{\hbar v} \Bigg\{ & G(2 p_0+q) \exp \left[ -\frac{\ii}{\hbar} (2 p_0+q) s \right] \\
+& G(2 p_0-q) \exp \left[ -\frac{\ii}{\hbar} (2 p_0-q) s \right] \Bigg\} . \nonumber
\end{align}

Solution~(\ref{Eq:Dirac_characteristic_spinor_solution}) is the main result in this section. It comprises the full (perturbative) effect of mass fluctuations on a massless Dirac particle propagating at initial momentum $p_0$, including disorder-induced state distortion, disorder-induced dephasing, disorder-induced backscattering, and disorder-induced Zitterbewegung. While the corresponding density matrix $\overline{\rho}(t)$ can be recovered by reversing the respective Fourier transforms, observables can in general also be determined from the characteristic function directly. We stress that the initial state $\vec{\chi}_0(s,q)$ is not further specified beyond the consistency requirement of being centered around $p_0$ in momentum space. Moreover, as one can easily check, the state is normalized at all times: $\chi_t^+(0,0)+\chi_t^-(0,0) = 1$.

For example, we now recover the disorder-induced backscattering, which, in the case of a Dirac particle, amounts to scattering among the two spin components. To this end, we evaluate the momentum distribution
\begin{align}
\vec{P}_t(p) \equiv& \left( \begin{array}{c} \bra{p} \overline{\rho}_{\uparrow \uparrow} \ket{p} \\ \bra{p} \overline{\rho}_{\downarrow \downarrow} \ket{p} \end{array} \right) = \int \dd x \left( \begin{array}{c} W_t^{+}(x,p) \\ W_t^{-}(x,p) \end{array} \right) \nonumber \\
=& \frac{1}{2 \pi \hbar} \int \dd s \, \ee^{-\frac{\ii}{\hbar} p s} \vec{\chi}_t(s,0) .
\end{align}
Note that, in the absence of disorder, $\vec{\chi}_t(s,q) = \exp(-\frac{\ii}{\hbar}v t q \sigma_z) \vec{\chi}_0(s,q)$, i.e., the momentum distribution is time-independent. In the presence of disorder, assuming a right-moving initial state, $\vec{\chi}_0(s,0) = \chi_0(s,0) (1,0)^T$, and since $\vec{\chi}_t(s,0) = \ee^{-\overline{F}_t^{(g)}(0,0) \mathbb{1}_2} \{ \mathbb{1}_2 \cosh[\overline{F}_t^{(g)}(s,0)] + \sigma_x \sinh[\overline{F}_t^{(g)}(s,0)] \} \vec{\chi}_0(s,0)$, we obtain
\begin{align} \label{Eq:Dirac_momentum_distribution}
\vec{P}_t(p) = \left( \begin{array}{c} \big\{ 1-\frac{2 \pi t}{\hbar v} G(2 p_0) \big\} P_0(p) \\ \frac{2 \pi t}{\hbar v} G(2 p_0) P_0(p+2 p_0) \end{array} \right) ,
\end{align}
with $P_0(p) = \frac{1}{2 \pi \hbar} \int \dd s \, \ee^{-\frac{\ii}{\hbar} p s} \chi_0(s,0)$ the momentum distribution of the initial state, centered around $p_0$. For instance, a Gaussian initial state of width $\sigma$ ($\hbar/\sigma \ll p_0$), $\psi_0(x) = \exp \left[ -\frac{1}{4} (\frac{x}{\sigma})^2 + \ii \frac{p_0 x}{\hbar} \right]/\sqrt{\sqrt{2 \pi} \sigma}$, comes with the characteristic function $\chi_0(s,q) = \exp \left[ -\frac{1}{8} (\frac{s}{\sigma})^2 - \frac{1}{2} (\frac{q \sigma}{\hbar})^2 + \ii \frac{p_0 s}{\hbar} \right]$ and the momentum distribution $P_0(p) = \sqrt{\frac{2}{\pi}} \frac{\sigma}{\hbar} \exp \left[ -\frac{2 \sigma^2}{\hbar^2} (p-p_0)^2 \right]$. To obtain (\ref{Eq:Dirac_momentum_distribution}), we used (\ref{Eq:Dirac_disorder_impact_large_time}) and assumed that $\frac{2 \pi t}{\hbar v} G(2 p_0) \ll 1$; the latter reflects our earlier assumption of weak disorder, i.e., within the temporal validity the disorder causes only a weak deviation from the unperturbed evolution.

Equation~(\ref{Eq:Dirac_momentum_distribution}) describes, within our approximation, the linear-in-time redistribution of the particle's state from right-moving centered around $p_0$ to left-moving centered around $-p_0$, cf.~Fig.~\ref{Fig:Dirac_backscattering}. We thus find that the disorder-dressed evolution recovers the backscattering of massless Dirac particles induced by mass fluctuations. Similarly to the case of a particle in a parabolic band and subject to potential/diagonal disorder, backscattering is controlled by the interplay between the disorder correlation length $\ell$ and the incident momentum $p_0$, mediated by the momentum transfer distribution $G(p)$ \cite{Gneiting2017quantum}. For instance, in the case of Gaussian correlations, $C(x) = C_0 \, \exp[-(x/\ell)^2]$, we obtain $G(q) = \frac{C_0 \ell}{2 \sqrt{\pi} \hbar} \exp \left[ -\left( q \ell/2 \hbar \right)^2 \right]$, which gives rise to exponentially suppressed backscattering if $p_0 \gg \hbar/\ell$.

In the backscattering-suppressed regime $p_0 \gg \hbar/\ell$, the evolution of the purity $r(t) = \Tr[\overline{\rho}(t)^2] = \frac{1}{2 \pi \hbar} \int \dd s \dd q \, [\chi_t^+(s,q) \chi_t^+(-s,-q) + \chi_t^-(s,q) \chi_t^-(-s,-q)]$ is then, in the limit $v t \gg \ell, \sigma$, approximated by $r(t) = 1- \frac{C_0}{v2 p_0^2} \left( 1- \ell/\sqrt{\ell^2 + 4 \sigma^2} \right)$, where we have assumed an initial Gaussian state with $p_0 \gg \hbar/\sigma$ and a small purity reduction. We thus find that the purity loss reaches a plateau, similarly to the purity loss induced by potential disorder \cite{Gneiting2017disorder}, and hence with similar implications for the transport of quantum information.

Finally, we derive the {\it disorder-induced Zitterbewegung}, i.e., disorder-induced oscillations of the position expectation value in the backscattering-suppressed regime $p_0 \gg \hbar/\ell$. Recall that unperturbed massless Dirac particles propagate linearly, with no exchange between right- and left-moving state components.

In principle, we could use solution (\ref{Eq:Dirac_characteristic_spinor_solution}) to evaluate the expectation value of the position operator. Here, we choose an alternative route, directly based on the right- and left-moving state components. As the right-(left-) moving state fraction is captured by $\chi_t^+(0,0)$ ($\chi_t^-(0,0)$), where $\chi_t^+(0,0) + \chi_t^-(0,0) = 1$, the position expectation value can be written as $\langle x \rangle (t) = \langle x \rangle_0 + \int_0^t d t' \{ \chi_{t'}^+(0,0) v - \chi_{t'}^-(0,0) v \}$. On the other hand, for a right-moving initial state $\chi_0^+(0,0) = 1$, we can infer from (\ref{Eq:Dirac_characteristic_spinor_solution}) that $\chi_t^+(0,0) = \frac{1}{2} \big( 1 + \exp[-2 \overline{F}_t^{(g)}(0,0)] \big)$. If we then use (\ref{Eq:Dirac_disorder_impact_even_component}) to approximate $\overline{F}_t^{(g)}(0,0)$ in the limit $p_0 \gg \hbar/\ell$ as $\overline{F}_t^{(g)}(0,0) = \frac{C_0}{p_0^2 v^2} \sin^2\left[ \frac{p_0 v t}{\hbar} \right]$, with $C_0 = \int d q \, G(q)$, we obtain, for weak disorder $C_0/p_0^2 v^2 \ll 1$,
\begin{align}
\langle x \rangle (t) = \langle x \rangle_0 + \left( v - \frac{C_0}{p_0^2 v} \right) t + \frac{C_0 \hbar}{2 v^2 p_0^3} \sin \left[ \frac{2 p_0 v t}{\hbar} \right] .
\end{align}
We thus find that mass perturbations induce a reduction of the average velocity, along with the signature oscillations of Zitterbewegung. This constitutes yet another example of how a time-resolved treatment of the disorder-averaged state reveals structural insights into the generic disorder impact. Let us note that, due to the disorder-independent frequency, these oscillations can be observed for individual disorder realizations (``quenched'' disorder), albeit with fluctuating amplitude. This disorder-induced Zitterbewegung may be directly observable in engineered platforms \cite{Keil2013random, Edmonds2012anderson, Maier2018environment, Arute2019quantum}. If and how the effect can be probed, possibly indirectly, e.g., in electronic systems, could be the subject of more targeted research.

\section{Conclusions}

Based on the coupled disorder channels ansatz, we derived the general disorder-dressed evolution equation (\ref{Eq:perturbative_master_equation}) for the disorder-averaged state, and demonstrated its application range with the two examples of a central spin in a spin bath and a random mass Dirac particle. In the first example, we described how the isotropic environment gives rise to state-dependent purity oscillations of purely incoherent nature. Such analysis may be instructive, e.g., to determine optimal readout times in quantum sensing or gate applications, minimizing the disorder impact. In the second example, featuring quantum transport, we recovered the backscattering induced by mass fluctuations, in a scenario where otherwise Klein tunneling reigns. Similarly, the disorder-induced {\it Zitterbewegung} is absent in unperturbed massless (one-dimensional) Dirac particles. Both examples demonstrated how the disorder-averaged evolution reflects the symmetries that are restored on the level of the collective behavior, i.e., rotational symmetry in the case of the central spin, and translational symmetry in the case of the Dirac particle.

Besides providing a comprehensive description of the perturbative disorder effect in a quantum optics and information language, this approach allows one to assess and quantify the disorder impact in terms of the coherence properties of the disorder-averaged state, a feature which is not reflected by averaged states in classical physics and which may help to identify disorder-robust system features, and ultimately, to design robust device architectures. On the other hand, engineered, highly controlled quantum systems are now used to experimentally explore disorder physics with unprecedented precision \cite{Maier2018environment, Delbecq2016quantum, Klimov2018fluctuations, Nakajima2018coherent, Arute2019quantum}, rendering it possible to experimentally test refined predictions on the level of the disorder-averaged quantum state.

To extend its scope of application, generalizing the framework, e.g., to time-dependent system Hamiltonians, and/or open systems appears desirable. This would not only make it possible to treat also more involved quantum control problems, but also give rise to a unified description of the two noise sources disorder and environment coupling. The coupled disorder channels (\ref{Eq:coupled_disorder_channels}) appear to be a suitable starting point for such generalizations.

\paragraph*{Acknowledgments.} The author thanks Alexander Rozhkov, Jurgen Smet, Daniel Leykam, Jung-Yun Han, Peter Stano, Daniel Burgarth, Henning Schomerus, and Jorge Puebla for helpful discussions, and Franco Nori for valuable comments on the manuscript.

\bibliography{literature}

\end{document}